\documentstyle[12pt]{article}

\begin{document}
\begin{flushright}
\baselineskip=12pt
UPR-969-T \\
IC/2001/164\\
ANL-HEP-PR-01-114\\
\end{flushright}

\begin{center}
\vglue 1.0cm
{\Large\bf N=2 6-dimensional Supersymmetric $E_6$ Breaking}
\vglue 2.0cm
{\Large Chao-Shang Huang$^{a}$, Jing Jiang$^{b}$, 
Tianjun Li$^{c}$~\footnote{E-mail: tli@bokchoy.hep.upenn.edu,
phone: (215) 573-5820, fax: (215) 898-2010.} and
 Wei Liao$^{d}$}
\vglue 1cm
{$^{a}$Institute of Theoretical Physics, Academia Sinica, P. O. Box
2735, \\
Beijing 100080, P. R. China\\
$^{b}$HEP Division, Argonne National Laboratory, \\
9700 S. Cass Avenue, Argonne, IL 60439\\
$^{c}$Department of Physics and Astronomy, 
University of Pennsylvania, \\
Philadelphia, PA 19104 \\
$^{d}$The Abdus Salam International Center for Theoretical Physics,\\
Strada Costiera 11, 34014 Trieste, Italy}
\end{center}

\vglue 1.0cm
\begin{abstract}
We study the $N=2$ supersymmetric $E_6$ models on the 
6-dimensional space-time where the supersymmetry and 
gauge symmetry can be broken by the discrete symmetry. 
On the space-time 
$M^4\times S^1/(Z_2\times Z_2') \times S^1/(Z_2\times Z_2')$,
for the zero modes, we obtain the 4-dimensional $N=1$
supersymmetric models with gauge groups 
$SU(3)\times SU(2) \times SU(2) \times U(1)^2$, 
$SU(4)\times SU(2) \times SU(2) \times U(1)$, and
$SU(3)\times SU(2) \times U(1)^3$ with one extra pair of Higgs doublets
from the vector multiplet.
In addition, considering that
the extra space manifold is the annulus $A^2$ and disc $D^2$,
we list all the constraints on constructing the 4-dimensional $N=1$
supersymmetric $SU(3)\times SU(2) \times U(1)^3$ models for the 
zero modes, and give the simplest model with $Z_9$ symmetry.
We also comment on the extra gauge symmetry breaking and 
its generalization.
\\[1ex]
\\[1ex]

\end{abstract}

\vspace{0.5cm}
\begin{flushleft}
\baselineskip=12pt
December 2001\\
\end{flushleft}
\newpage
\setcounter{page}{1}
\pagestyle{plain}
\baselineskip=14pt

\section{Introduction}
Grand Unified Theory (GUT) gives us an simple
 and elegant understanding of the quantum numbers of quarks and leptons,
and the success of gauge coupling unification in the Minimal
Supersymmetric
Standard Model strongly supports
 this idea. Although the Grand Unified Theory at high energy scale has
been widely accepted now, there are some problems in GUT: 
 the grand unified
gauge symmetry breaking mechanism, the doublet-triplet splitting problem,
and
the proton decay, etc.

Recently, a new scenario proposed to address above questions in GUT has
 been discussed extensively~\cite{SBPL, JUNI, JUNII}.
 The key point is that the GUT
gauge symmetry exists in 5 or higher dimensions and is broken down to the
4-dimensional 
$N=1$ supersymmetric Standard Model like gauge symmetry for
 the zero modes due to the 
discrete symmetries in the neighborhoods of the branes
or on the extra space manifolds, which
become non-trivial constraints on the multiplets and gauge generators in 
GUT~\cite{JUNII}. 
The attractive models have been constructed explicitly, where
the supersymmetric 5-dimensional and 6-dimensional 
GUT models are broken down to
the 4-dimensional $N=1$
 supersymmetric $SU(3)\times SU(2) \times U(1)^{n-3}$
model, where $n$ is the rank of GUT group, through the 
compactification on various orbifolds and manifolds. 
The GUT gauge symmetry breaking and doublet-triplet
splitting problems have been solved neatly by the
discrete symmetry projections.
Other interesting phenomenology, like $\mu$ problems, gauge coupling
unifications, non-supersymmetric GUT, gauge-Higgs unification,
proton decay, etc, have also been discussed~\cite{SBPL, JUNI, JUNII}. 

All of the models~\cite{SBPL, JUNI, JUNII} discussed previously
have gauge group $SU(N)$ or $SO(N)$. So, we study the
$E_6$ model in the present paper, which is as interesting as
 $SU(5)$ and $SO(10)$ GUT models. Because $E_6$ is a rank 6 exceptional
group, in order to break the gauge symmetry and supersymmetry, we need to
consider at least two extra dimensions. In addition, the 6-dimensional 
$N=1$ supersymmetric theory is chiral, where the gaugino (and gravitino) 
has positive chirality and the matters (hypermultiplets) have negative
chirality, 
so, it often has anomaly unless we put the
Standard Model fermions on the brane, and add one multiplet in
the adjoint representation of the gauge group or some suitable
matter contents in the bulk to cancel the 
gauge anomaly. And the 6-dimensional non-supersymmetric $E_6$ models and
$N=1$ 
supersymmetric $E_6$ models can be considered as special cases of
$N=2$ supersymmetric $E_6$ models,
therefore, we only discuss the 6-dimensional $N=2$ supersymmetric $E_6$
models.
Moreover,
because $N=2$ 6-dimensional supersymmetric theory has 16 real
supercharges,
which corresponds to $N=4$ 4-dimensional supersymmetric theory,
we can not have hypermultiplets in the bulk. Therefore,
we have to put the Standard Model fermions on the brane or brane
intersection.

In this paper, we first review the discussions of $E_6$ 
breaking by Wilson line
in our context~\cite{Wilson}. Then, we study $E_6$ breaking
on the space-time $M^4\times S^1/(Z_2\times Z_2') \times S^1/(Z_2\times
Z_2')$,
where $M^4$ is the 4-dimensional Minkowsky space-time.
For the zero modes, we obtain the 4-dimensional $N=1$
supersymmetric models with gauge groups 
$SU(3)\times SU(2) \times SU(2) \times U(1)^2$, 
$SU(4)\times SU(2) \times SU(2) \times U(1)$, and
$SU(3)\times SU(2) \times U(1)^3$ with one extra pair of Higgs doublets
from the vector multiplet.
In addition, considering that
the extra space manifold is an annulus $A^2$ and a disc $D^2$,
we can define $Z_n$ symmetry on the extra space manifold.
We list all the constraints on constructing
 the 4-dimensional $N=1$
supersymmetric $SU(3)\times SU(2) \times U(1)^3$ model for the zero modes,
and give the simplest model with $Z_9$ symmetry.
Furthermore, we comment on the extra gauge symmetry breaking and 
its generalization.       

We would like to explain our convention. For simplicity, we define 
\begin{equation}
  (\alpha, \beta, \gamma) \equiv \left(\begin{array}{ccc}
    \alpha & 0 & 0  \\ 
    0 & \beta & 0 \\ 
    0 & 0 & \gamma \\
  \end{array} \right)~.~\,
\end{equation}
In addition, suppose $G$ is a Lie group and $H$ is a subgoup of $G$.
In general, for $G=SU(N)$ and $G=SO(N)$, $H$ can be
the subgroup of $U(N)$ and $O(N)$, respectively.
We denote the commutant of $H$ in $G$ as $G/H$, {\it i. e.},
\begin{equation}
G/H\equiv \{g \in G|gh=hg, ~{\rm for ~any} ~ h \in H\}~.~\,
\end{equation}
And if $H_1$ and $H_2$ are the subgroups of $G$,
we define
\begin{equation}
G/\{H_1 \cup H_2\}\equiv \{G/H_1\} \cap \{G/H_2\}~.~\,
\end{equation}

\section{Background of $E_6$ Breaking}
In 1985, a lot of work has been done on
$E_6$ breaking because
the $E_6$ model can be obtained from the compactification
of the weakly coupled heterotic
$E_8\times E_8$ string theory on the Calabi-Yau manifold
by spin connection embedding~\cite{Wilson, Esix}. 
We would like to review the discussions 
of $E_6$ breaking by Wilson line in our context~\cite{Wilson}, which will
be 
used to discuss $E_6$ breaking in this paper.

Suppose the space-time manifold is $M^4\times K$ where $K$ is the
$k$-dimensional
extra space manifold, and we can define a discrete symmetry $\Gamma$ on
$K$.
In general, $\Gamma$ can be the product of discrete groups, 
and $\Gamma$ may not act freely on $K$. And when it does not act freely on
$K$,
there exists a brane at 
each fixed point, line or hypersurface, where the Standard Model fermions
can be located. 

The gauge fields of $E_6$ are in the adjoint representation 
of $E_6$ with dimension {\bf 78}, and
$E_6$ contains a maximal subgroup $SU(3)_C \times SU(3)_L \times
SU(3)_R$ where $SU(3)_C$ is color $SU(3)$, 
$SU(3)_L$ and $SU(3)_R$ describe the
weak interactions of left-handed and right-handed quarks,
respectively~\cite{Group}.
 Under the gauge groups $SU(3)_C \times SU(3)_L \times
SU(3)_R$, the $E_6$ gauge fields decompose to {\bf (8, 1, 1), (1, 8, 1), 
(1, 1, 8), (${\bf \bar 3}$, 3, 3), (3, ${\bf \bar 3, \bar
3}$)}~\cite{Group}.

For simplicity,
we assume that $\Gamma$ is $Z_n$ in the following discussions, where 
$Z_n$ is generated by the n-th roots of unity.
Since the discussions for the product of cyclic groups are similar, 
 we do not repeat them here. Let $\gamma$ be a generator of
$\Gamma$, we choose the following matrix representation
for $\gamma$, which will give us the representations of all the elements
in
$\Gamma$,
\begin{equation}
R_{\gamma}=(+1, +1, +1) \otimes (\alpha, \alpha, \beta)
\otimes (\delta, \rho, \sigma)~,~\,
\end{equation}
where $\alpha^n=\beta^n=\delta^n=\rho^n=\sigma^n=1$. 
And the map $R:~ \Gamma \longrightarrow R_{\Gamma} \subset
SU(3) \times (SU(3)\times Z_n) \times (SU(3)\times Z_n)
\subset SU(3)\times U(3)\times U(3)$ must
be a homomorphism. 

We would like to make a remark here. If we discussed the $SU(N)$ 
 breaking on the extra space-manifold with 
$Z_n$ symmetry, we can choose the representation of $Z_n$ in 
$SU(N)\times Z_n \subset U(N)$. And if we discussed the $SO(N)$ breaking 
on the extra space-manifold with $Z_n$ symmetry, 
we can choose the representation of $Z_n$ in 
$SO(N)$ if $n$ is odd and in $SO(N)\times Z_2 \approx O(N)$ if $n$ is
even.
The same rule applies for the product of cyclic groups.
However, we are interested in $E_6$ breaking, not 
$SU(3)_C \times SU(3)_L \times SU(3)_R$ breaking in this paper.
For an arbitrarily choice of $R_{\gamma}$, 
the commutant of $R_{\Gamma}$ in $E_6$ 
($E_6/R_{\Gamma}$) might not form a group if 
$R_{\gamma}$ were not a subgoup of $E_6$. Because we require 
 $E_6/R_{\Gamma}$ to be a group, we choose that 
$R_{\Gamma} \subset SU(3)_C \times SU(3)_L \times SU(3)_R$ in
the following discussions, i. e., $\alpha^2 \beta = \delta \rho \sigma=1$.
With the choice of $R_{\Gamma} \subset SU(3)_C \times SU(3)_L \times
SU(3)_R$, 
we can embed all the generators back to those of $E_6$. 
Because $R_{\Gamma}$ is an abelian subgroup of $E_6$, it is easy to 
prove that $E_6/R_{\Gamma}$ also forms a subgroup of $E_6$ with rank 6.
However, there is an exception in our discussions. 
For $Z_2$ case, in order to break $E_6$ down to 
$SU(3)_C \times SU(3)_L \times SU(3)_R$, we 
choose $R=(+1, +1, +1)  \otimes (+1, +1, +1)
\otimes (-1, -1, -1)$ or $R=(+1, +1, +1)  \otimes (-1, -1, -1)
\otimes (+1, +1, +1)$.

Now, we discuss the $E_6$
breaking. For simplicity, we just consider $E_6$ gauge field 
$A_{\mu}=A_{\mu}^{B} T^{B}$, 
where $\mu=0, 1, 2, 3$ and $B=1, 2, ..., 78$. Let us denote the
$E_6$ gauge fields in $SU(3)_C \times SU(3)_L \times
SU(3)_R$ as $A_{\mu}=A_{\mu}^{a} T^a$, where $T^a$ is 
the product of three $3\times 3$ 
matrices, for example, if $t^b$ was a Lie algebra for $SU(3)_C$, the
corresponding $T^b=t^b \otimes (+1, +1, +1)
\otimes (+1, +1, +1)$. We also denote the $E_6$ gauge fields in 
${\bf (\bar 3, 3, 3)}$ and
 ${\bf (3, \bar 3, \bar 3)}$ as  $A_{\mu}=A_{\mu}^{\hat a} T^{\hat a}$
and $A_{\mu}=A_{\mu}^{\bar a} T^{\bar a}$, respectively,
 where $T^{\hat a}$ and $T^{\bar a}$ are
the products of three $3\times 1$ columns.

Because the extra space manifold has $\Gamma=Z_n$ symmetry,  for any 
$\gamma \in \Gamma$, we have
\begin{eqnarray}
A_{\mu}^B (x^{\mu}, \gamma_i y^1, \gamma_i y^2, ..., \gamma_i y^k) T^{B} = 
(R_{\gamma})^{l_A} A_{\mu}^B (x^{\mu},  y^1, y^2, ..., y^k) T^{B}
(R_{\gamma}^{-1})^{m_A}
~,~\,
\end{eqnarray}
where $y^i$ for $i=1, 2, ..., k$ are the coordinates for the
extra space manifold, and $(l_A, m_A)$ are equal to
$(1, 1)$, $(1,0)$, and $(0, 1)$ for $B=a, \hat a, \bar a,$ respectively.
The $E_6$ gauge fields $A_{\mu}^B$ will have zero modes only if
\begin{eqnarray}
A_{\mu}^B (x^{\mu}, \gamma_i y^1, \gamma_i y^2, ..., \gamma_i y^k) =
 A_{\mu}^B (x^{\mu},  y^1, y^2, ..., y^k)~.~\,
\end{eqnarray}
The zero modes of gauge fields form the group $E_6/R_{\Gamma}$ with
rank 6.

From the phenomenological point of view~\cite{Wilson}, 
for the zero modes, we require that:
(1) $SU(3)_L$ and $SU(3)_R$ can not be completely
unbroken for otherwise their couplings would evolve at low energy to be as
strong as that of $SU(3)_C$; (2) To avoid the proton decay,
the unbroken subgroups do not contain $SU(5)$, $SU(6)$ and $SO(10)$.
The first requirement implies that $\alpha \not= \beta$, and that
$\delta$, $\rho$, $\sigma$ are not all equal.
The second requirement implies that we can have at most 
one pair of color triplets.

As an example, if $\delta \not= \rho \not= \sigma$, and
$\alpha \delta, \alpha \rho, \alpha \sigma, 
\beta \delta, \beta \rho, \beta \sigma$ are all not equal to 
one, for the zero modes, the gauge group is
$SU(3)_C \times SU(2)_L \times U(1)^3$. And for $Z_2$, if
 we chose $R=(+1, +1, +1)  \otimes (+1, +1, +1)
\otimes (-1, -1, -1)$ or $R=(+1, +1, +1)  \otimes (-1, -1, -1)
\otimes (+1, +1, +1)$, we break $E_6$ down to
$SU(3)_C \times SU(3)_L \times SU(3)_R$.

\section{$E_6$ Breaking on $M^4\times S^1/(Z_2\times Z_2') \times
S^1/(Z_2\times Z_2')$}
In this section, we will discuss $E_6$ breaking on 
the space time $M^4\times S^1/(Z_2\times Z_2') \times S^1/(Z_2\times
Z_2')$.
We consider 
the 6-dimensional space-time which can be factorized into the product of
the 
ordinary 4-dimensional Minkowski space-time $M^4$, and the torus $T^2$
which is homeomorphic to $S^1\times S^1$. The corresponding
coordinates for the space-time are $x^{\mu}$, ($\mu = 0, 1, 2, 3$),
$y\equiv x^5$ and $z\equiv x^6$. 
And the radii for the circles along the $y$ direction and $z$ direction
are
$R_1$ and $R_2$, respectively.
We also define $y'$ and $z'$ by $y' \equiv y-\pi R_1/2$ 
and $z' \equiv z- \pi R_2/2$. The orbifold 
$S^1/(Z_2\times Z_2') \times S^1/(Z_2\times Z_2')$
is obtained from $S^1\times S^1$
by moduloing the following equivalent classes:
\begin{equation}
 y\sim -y~,~ z\sim -z~,~ y'\sim -y'~,~ z'\sim -z'~.~\,
\end{equation}
The cooresponding operators for the $Z_2$ symmetries
 $y\sim -y,~ z\sim -z,~ y'\sim -y'$ and $z'\sim -z'$ are 
$P^y$, $P^z$, $P^{y'}$ and $P^{z'}$, respectively.
Allowing a little abuse of notation, we also denote
the matrix representations of $P^y$, $P^z$, $P^{y'}$ and $P^{z'}$
as $P^y$, $P^z$, $P^{y'}$ and $P^{z'}$, as used in the
literature~\cite{SBPL, JUNI, JUNII}.

Let us explain the 6-dimensional gauge theory with $N=2$ supersymmetry.
$N=2$ supersymmetric theory in 6-dimension has 16 real supercharges,
corresponding to $N=4$ supersymmetry in 4-dimension. Therefore, only the
vector multiplet can be introduced in the bulk, and 
 the Standard Model fermions are confined on the 4-branes, 3-branes or
4-brane intersections. In terms of the 4-dimensional
$N=1$ supersymmetry language, 
the theory contains a vector multiplet $V(A_{\mu}, \lambda_1)$
in which $\lambda_1$ is the gaugino,
and three chiral multiplets $\Sigma_5$, $\Sigma_6$, and $\Phi$. All 
of them are in the adjoint representation of the gauge group. In addition,
the $\Sigma_5$ and $\Sigma_6$ chiral multiplets
contain the gauge fields $A_5$ and $A_6$ in
their lowest components, respectively.

In the Wess-Zumino gauge and 4-dimensional $N=1$ 
supersymmetry language, the bulk action 
is~\cite{NAHGW}
\begin{eqnarray}
  S &=& \int d^6 x \Biggl\{
  {\rm Tr} \Biggl[ \int d^2\theta \left( \frac{1}{4 k g^2} 
  {\cal W}^\alpha {\cal W}_\alpha + \frac{1}{k g^2} 
  \left( \Phi \partial_5 \Sigma_6 - \Phi \partial_6 \Sigma_5
  - \frac{1}{\sqrt{2}} \Phi 
  [\Sigma_5, \Sigma_6] \right) \right) 
\nonumber\\
&& + {\rm H.C.} \Biggr] 
  + \int d^4\theta \frac{1}{k g^2} {\rm Tr} \Biggl[ 
  \sum_{i=5}^6 \left((\sqrt{2} \partial_i + \Sigma_i^\dagger) e^{-V} 
  (-\sqrt{2} \partial_i + \Sigma_i) e^{V} + 
   \partial_i e^{-V} \partial_i e^{V}\right)
\nonumber\\
  && \qquad \qquad \qquad
  + \Phi^\dagger e^{-V} \Phi e^{V}  \Biggr] \Biggr\} ~.~\,
\label{eq:5daction}
\end{eqnarray}
And the gauge transformation is given by
\begin{eqnarray}
  e^V &\rightarrow& e^\Lambda 
    e^V e^{\Lambda^\dagger}, \\
  \Sigma_i &\rightarrow& e^\Lambda (\Sigma_i - \sqrt{2} \partial_i) 
    e^{-\Lambda}, \\
  \Phi &\rightarrow& e^\Lambda \Phi e^{-\Lambda}~,~\,
\end{eqnarray}
where $i=5, 6$.

From the action, we obtain 
the transformations of vector multiplet
 under the $Z_2$ operators $P^y$,
$P^z$
\begin{eqnarray}
  V(x^{\mu},-y, z) &=& (P^y)^{l_V} V(x^{\mu}, y, z) ((P^y)^{-1})^{m_V}
~,~\,
\end{eqnarray}
\begin{eqnarray}
  \Sigma_5(x^{\mu},-y, z) &=& - (P^y)^{l_{\Sigma_5}} \Sigma_5(x^{\mu}, y,
z) 
((P^y)^{-1})^{m_{\Sigma_5}}
~,~\,
\end{eqnarray}
\begin{eqnarray}
  \Sigma_6(x^{\mu},-y, z) &=&  (P^y)^{l_{\Sigma_6}} \Sigma_6(x^{\mu}, y,
z) 
((P^y)^{-1})^{m_{\Sigma_6}}
~,~\,
\end{eqnarray}
\begin{eqnarray}
  \Phi(x^{\mu},-y, z) &=& - (P^y)^{l_{\Phi}} \Phi(x^{\mu}, y, z) 
((P^y)^{-1})^{m_{\Phi}}
~,~\,
\end{eqnarray}
\begin{eqnarray}
  V(x^{\mu},y, -z) &=& (P^z)^{l_V} V(x^{\mu}, y, z) ((P^z)^{-1})^{m_V}
~,~\,
\end{eqnarray}
\begin{eqnarray}
  \Sigma_5(x^{\mu}, y, -z) &=&  (P^z)^{l_{\Sigma_5}}
 \Sigma_5(x^{\mu}, y, z) ((P^z)^{-1})^{m_{\Sigma_5}}
~,~\,
\end{eqnarray}
\begin{eqnarray}
  \Sigma_6(x^{\mu}, y, -z) &=&  - (P^z)^{l_{\Sigma_6}}
 \Sigma_6(x^{\mu}, y, z) ((P^z)^{-1})^{m_{\Sigma_6}}
~,~\,
\end{eqnarray}
\begin{eqnarray}
  \Phi(x^{\mu}, y, -z) &=& - (P^z)^{l_{\Phi}} \Phi(x^{\mu}, y, z) 
((P^z)^{-1})^{m_{\Phi}}
~,~\,
\end{eqnarray}
where $(l_V, m_V)$,  $(l_{\Sigma_5}, m_{\Sigma_5})$,
$(l_{\Sigma_6}, m_{\Sigma_6})$, and $(l_{\Phi}, m_{\Phi})$  
 are equal to $(1, 1)$  if the gauge fields were
in the representations ${\bf (8, 1, 1), (1, 8, 1), 
(1, 1, 8)}$, and
$(l_V, m_V)$,  $(l_{\Sigma_5}, m_{\Sigma_5})$,
$(l_{\Sigma_6}, m_{\Sigma_6})$, and $(l_{\Phi}, m_{\Phi})$  
 are equal to
$(1, 0)$ if the gauge fields were
in the representation ${\bf (\bar 3, 3, 3)}$,
and $(l_V, m_V)$,  $(l_{\Sigma_5}, m_{\Sigma_5})$,
$(l_{\Sigma_6}, m_{\Sigma_6})$, and $(l_{\Phi}, m_{\Phi})$  
 are equal to
$(0, 1)$ if the gauge fields were
in the representation ${\bf (3, \bar 3, \bar 3)}$. Moreover,
the transformations of
 vector multiplet under the $Z_2$ operators $P^{y'}$ and
$P^{z'}$ are similar.

In the following models, for the zero modes,
we will break the 4-dimensional $N=4$
supersymmetry down to $N=1$ supersymmetry, and break the $E_6$ gauge group
down to $E_6/\{P^y \cup P^z \cup P^{y'} \cup P^{z'}\}$.
Including the KK modes, the intersection 3-branes and boundary 4-branes
preserve
the 4-dimensional $N=1$ and $N=2$ supersymmetry, respectively.
The general 4-dimensional supersymmetry and gauge groups on  
the intersection 3-branes and boundary 4-branes are given in the Table 1.
The KK mode expansions and the detail of this set-up 
can be found in Ref.~\cite{JUNI}.

\renewcommand{\arraystretch}{1.4}
\begin{table}[t]
\caption{For $E_6$ models on $S^1/(Z_2\times Z_2') \times S^1/(Z_2\times
Z_2')$, 
 the number of 4-dimensional supersymmetry and gauge groups on the
3-branes, which
are located at the fixed points $(y=0, z=0),$ $(y=0, z=\pi R_2/2),$ 
$(y=\pi R_1/2, z=0)$, and 
$(y=\pi R_1/2, z=\pi R_2/2)$, or on the 4-branes which are located at the
fixed lines
$y=0$, $z=0$, $y=\pi R_1/2$, $z=\pi R_2/2$.
\label{tab:SUV11}}
\vspace{0.4cm}
\begin{center}
\begin{tabular}{|c|c|c|}
\hline        
Brane Position & SUSY & Gauge Symmetry\\ 
\hline
$(0, 0) $ & $N=1$ & $G/\{P^y\cup P^z\}$ \\
\hline
$(0, \pi R_2/2)$  & $N=1$ & $G/\{P^y\cup P^{z'}\}$ \\
\hline
$(\pi R_1/2, 0) $  & $N=1$ & $G/\{P^{y'}\cup P^{z}\} $ \\
\hline
$(\pi R_1/2, \pi R_2/2) $ & $N=1$ & $G/\{P^{y'}\cup P^{z'}\} $ \\
\hline
$y=0$   & $N=2$ & $G/P^y$ \\
\hline
$z= 0 $   & $N=2$ & $G/P^z$ \\
\hline
$y=\pi R_1/2 $  & $N=2$ & $G/P^{y'}$ \\
\hline
$z=\pi R_2/2 $ & $N=2$ & $G/P^{z'}$ \\
\hline
\end{tabular}
\end{center}
\end{table}

\subsection{Models without the Zero Modes of $\Sigma_5$, $\Sigma_6$ and
$\Phi$}
We will first discuss the models without the zero modes of $\Sigma_5$,
$\Sigma_6$ and
$\Phi$. In order to project out all
the zero modes of $\Sigma_5$, $\Sigma_6$ and
$\Phi$, we choose the matrix representations of $P^z$ and $P^z$ as product
of three $3\times 3$ unit matrices
\begin{equation}
P^y=P^z=(+1, +1, +1)  \otimes (+1, +1, +1) \otimes (+1, +1, +1)
~.~\,
\end{equation}
So, considering the zero modes,
under $P^y$ projection, we can break the 4-dimensional $N=4$ supersymmetry
down to the
$N=2$ supersymmetry with $(V, \Sigma_6)$ forming a vector multiplet and
$(\Sigma_5, \Phi)$ forming a hypermultiplet, and we can break the
4-dimensional $N=2$ supersymmetry down to the $N=1$ supersymmetry further
by
the $P^z$ projection. 

We define 5 matrices which
will be used in the following discusssions
\begin{equation}
A=(+1, +1, +1)  \otimes (-1, -1, +1) \otimes (-1, -1, +1)
~,~\,
\end{equation}
\begin{equation}
B=(+1, +1, +1)  \otimes (+1, +1, +1) \otimes (-1, -1, +1)
~,~\,
\end{equation}
\begin{equation}
C=(+1, +1, +1)  \otimes (-1, -1, +1) \otimes (+1, +1, +1)
~,~\,
\end{equation}
\begin{equation}
D=(+1, +1, +1)  \otimes (+1, +1, +1) \otimes (-1, -1, -1)
~,~\,
\end{equation}
\begin{equation}
E=(+1, +1, +1)  \otimes (-1, -1, -1) \otimes (+1, +1, +1)
~.~\,
\end{equation}
Because $A, B, C, D, E$ are order 2 elements and the unit element (or
indentity)
$e$ commutes with all the elements in the group, we define
$E_6/A\equiv E_6/\{e, A\}$ for simplicity and similarly for the others.
As an example, we will explain how to obtain $E_6/A$. As we know, $E_6$
has three maximal subgroups with rank 6: $SO(10)\times U(1)$, 
$SU(6)\times SU(2)$ and $SU(3)\times SU(3)\times SU(3)$~\cite{Group}. 
Because $A$ 
 is an order 2 subgroup of $SU(3)\times SU(3)\times SU(3)$,
 $E_6/A$ forms a maximal subgroup and must be one of the three $E_6$
maximal subgroups with rank 6. $E_6/A$ has 46 gauge generators by simple
counting, therefore, we obtain that $E_6/A$ is $SO(10)\times U(1)$.
Similarly, we can calculate the other commutants.
In short, we have
\begin{equation}
E_6/A \approx SO(10)\times U(1)
~,~\,
\end{equation}
\begin{equation}
E_6/B \approx E_6/C \approx SU(6) \times SU(2)
~,~\,
\end{equation}
\begin{equation}
E_6/D \approx E_6/E \approx SU(3)\times SU(3)\times SU(3)
~,~\,
\end{equation}
\begin{equation}
E_6/\{A \cup B\} \approx E_6/\{A \cup C\} \approx E_6/\{B \cup C\} \approx
SU(4)\times SU(2) \times SU(2) \times U(1)
~,~\,
\end{equation}
\begin{equation}
E_6/\{A \cup D\} \approx E_6/\{A \cup E\} \approx
SU(3)\times SU(2) \times SU(2) \times U(1)^2~.~\,
\end{equation}

{\bf Model I.} We choose the matrix representations for $P^{y'}$ and
$P^{z'}$ as
\begin{equation}
P^{y'}=A~,~ P^{z'}=B~.~\,
\end{equation}
For the zero modes, the bulk 4-dimensional 
$N=4$ supersymmetric $E_6$ model is broken down to the
$N=1$ supersymmetric $SU(4)\times SU(2) \times SU(2) \times U(1)$ model.
Including the KK modes, 
the gauge groups on the intersection 3-branes at
$(y=0, z=0)$, $(y=0, z=\pi R_2/2)$, $(y=\pi R_1/2, z=0)$
and $(y=\pi R_1/2, z=\pi R_2/2)$ are
$E_6$, $SU(6) \times SU(2)$, $SO(10)\times U(1)$ and
$SU(4)\times SU(2) \times SU(2) \times U(1)$, respectively.
And the gauge groups on the 4-branes at $y=0$,
$z=0$, $y=\pi R_1/2$ and $z=\pi R_2/2$ are 
$E_6$, $E_6$, $SO(10)\times U(1)$ and $SU(6) \times SU(2)$, respectively.
Similarly, one can discuss the model by choosing $P^{y'}=A$ and
$P^{z'}=C$.

{\bf Model II.} We choose the matrix representations for $P^{y'}$ and
$P^{z'}$ as
\begin{equation}
P^{y'}=B~,~ P^{z'}=C~.~\,
\end{equation}
For the zero modes, the bulk 4-dimensional 
$N=4$ supersymmetric $E_6$ model is broken down to the
$N=1$ supersymmetric $SU(4)\times SU(2) \times SU(2) \times U(1)$ model.
Including the KK modes, 
the gauge groups on the intersection 3-branes at
$(y=0, z=0)$, $(y=0, z=\pi R_2/2)$, $(y=\pi R_1/2, z=0)$
and $(y=\pi R_1/2, z=\pi R_2/2)$ are
$E_6$, $SU(6) \times SU(2)$, $SU(6) \times SU(2)$ and
$SU(4)\times SU(2) \times SU(2) \times U(1)$, respectively.
And the gauge groups on the 4-branes at $y=0$,
$z=0$, $y=\pi R_1/2$ and $z=\pi R_2/2$ are 
$E_6$, $E_6$, $SU(6) \times SU(2)$ and $SU(6) \times SU(2)$, respectively.

{\bf Model III.} We choose the matrix representations for $P^{y'}$ and
$P^{z'}$ as
\begin{equation}
P^{y'}=A~,~ P^{z'}=D~.~\,
\end{equation}
For the zero modes, the bulk 4-dimensional 
$N=4$ supersymmetric $E_6$ model is broken down to the
$N=1$ supersymmetric $SU(3)\times SU(2) \times SU(2) \times U(1)^2$ model.
Including the KK modes, 
the gauge groups on the intersection 3-branes at
$(y=0, z=0)$, $(y=0, z=\pi R_2/2)$, $(y=\pi R_1/2, z=0)$
and $(y=\pi R_1/2, z=\pi R_2/2)$ are
$E_6$, $SU(3)\times SU(3)\times SU(3)$, $SO(10)\times U(1)$ and
$SU(3)\times SU(2) \times SU(2) \times U(1)^2$, respectively.
And the gauge groups on the 4-branes at $y=0$,
$z=0$, $y=\pi R_1/2$ and $z=\pi R_2/2$ are 
$E_6$, $E_6$, $SO(10)\times U(1)$ and $SU(3)\times SU(3)\times SU(3)$,
respectively.
Similarly, one can discuss the model by choosing $P^{y'}=A$ and
$P^{z'}=E$.

\subsection{Model with Gauge-Higgs Unification}
In this subsection, we will present the model with $SU(3)\times
SU(2)\times U(1)^3$ 
gauge symmetry and one pair of $SU(2)_L$ Higgs doublets from $\Phi$.
 We choose the matrix representations for $P^y$, $P^z$, $P^{y'}$ and
$P^{z'}$ as
\begin{equation}
P^y=P^z=A=(+1, +1, +1)  \otimes (-1, -1, +1) \otimes (-1, -1, +1)
~,~\,
\end{equation}
\begin{equation}
P^{y'}=(+1, +1, +1)  \otimes (-1, -1, +1) \otimes (-1, +1, -1)
~,~\,
\end{equation}
\begin{equation}
P^{z'}=(+1, +1, +1)  \otimes (-1, -1, +1) \otimes (+1, -1, -1)
~.~\,
\end{equation}
And we would like to point out the commutant groups
\begin{equation}
E_6/A \approx E_6/P^{y'} \approx E_6/P^{z'} \approx SO(10)\times U(1)
~,~\,
\end{equation}
\begin{equation}
E_6/\{A \cup P^{y'}\} \approx E_6/\{A \cup P^{z'}\} 
\approx E_6/\{P^{y'} \cup P^{z'}\} \approx
SU(5) \times U(1)^2
~,~\,
\end{equation}
\begin{equation}
E_6/\{A \cup P^{y'} \cup P^{z'}\} \approx SU(3)\times SU(2)\times
U(1)^3~.~\,
\end{equation}

We project out all the zero modes of $\Sigma_5$ and $\Sigma_6$ by
choosing $P^y=P^z$. And we project out all the zero modes of $\Phi$
except one pair of $SU(2)_L$ doublets, which can be viewed as one
pair of Higgs doublets. Considering the zero modes, the bulk 4-dimensional 
$N=4$ supersymmetric $E_6$ model is broken down to the
$N=1$ supersymmetric $SU(3)\times SU(2) \times U(1)^3$ model.
Including the KK modes, 
the gauge groups on the intersection 3-branes at
$(y=0, z=0)$, $(y=0, z=\pi R_2/2)$, $(y=\pi R_1/2, z=0)$
and $(y=\pi R_1/2, z=\pi R_2/2)$ are
 $SO(10)\times U(1)$, $SU(5) \times U(1)^2$, $SU(5) \times U(1)^2$,
and $SU(5) \times U(1)^2$, respectively.
In addition, the gauge groups on the 4-branes at $y=0$,
$z=0$, $y=\pi R_1/2$ and $z=\pi R_2/2$ are all 
 $SO(10)\times U(1)$.

\section{$E_6$ Breaking on $M^4\times A^2$ and $M^4\times D^2$}
In this section, we would like to discuss $E_6$ breaking 
on the space-time $M^4\times A^2$ and $M^4\times D^2$, where
$A^2$ and $D^2$ are the two dimensional annulus and disc, repectively.
And we only show the models with $SU(3)\times SU(2) \times U(1)^3$
gauge symmetry and 4-dimensional $N=1$ supersymmetry
for the zero modes. Similarly, one can discuss the
models with gauge groups $SU(3)\times SU(2) \times SU(2) \times U(1)^2$,
or $SU(4)\times SU(2) \times U(1)^2$, or 
$SU(4)\times SU(2) \times SU(2) \times U(1)$ 
and  4-dimensional $N=1$ supersymmetry for the zero modes.

The convenient coordinates for the annulus $A^2$ 
is polar coordinates $(r, \theta)$,
and it is easy to change them to the complex coordinates by $z=r e^{{\rm
i}\theta}$.
We call the innner radius of the annulus as $R_1$, and
the outer radius of the annulus as $R_2$. When $R_1=0$, the annulus
becomes
 the disc $D^2$, which is an special case of $A^2$. 
We can define the $Z_n$ symmetry on the annulus $A^2$ by the
equivalent class
\begin{eqnarray}
z\sim \omega z~,~\,
\end{eqnarray} 
where $\omega=e^{{\rm i} {{2 \pi}\over n}}$. And we denote
the corresponding generator for $Z_n$ as $\Omega$ which satifies
$\Omega^n=1$. 
The KK mode expansions and the detail of this set-up can be
found in Ref.~\cite{JUNII}.

The $N=2$ supersymmetry in 6-dimension
 corresponds to the $N=4$ supersymmetry in 4-dimension,
thus, only the gauge multiplet can be introduced in the bulk.  This
multiplet can be decomposed under the 4-dimensional
 $N=1$ supersymmetry into a vector
multiplet $V$ and three chiral multiplets $\Sigma$, $\Phi$, and $\Phi^c$
in the adjoint representation, with the fifth and sixth components of the
gauge
field, $A_5$ and $A_6$, contained in the lowest component of $\Sigma$.
The Standard Model fermions are on the boundary 4-brane at $r=R_1$
or $r=R_2$ for the annulus $A^2$ scenario, 
and on the 3-brane at origin or on the boundary 4-brane at $r=R_2$ for 
the disc $D^2$ scenario. 

In the Wess-Zumino gauge and 4-dimensional $N=1$ supersymmetry
language, the bulk action 
is~\cite{NAHGW}
\begin{eqnarray}
  S &=& \int d^6 x \Biggl\{
  {\rm Tr} \Biggl[ \int d^2\theta \left( \frac{1}{4 k g^2} 
  {\cal W}^\alpha {\cal W}_\alpha + \frac{1}{k g^2} 
  \left( \Phi^c \partial \Phi   - \frac{1}{\sqrt{2}} \Sigma 
  [\Phi, \Phi^c] \right) \right) + {\rm h.c.} \Biggr] 
\nonumber\\
  && + \int d^4\theta \frac{1}{k g^2} {\rm Tr} \Biggl[ 
  (\sqrt{2} \partial^\dagger + \Sigma^\dagger) e^{-V} 
  (-\sqrt{2} \partial + \Sigma) e^{V}\Biggr]
\nonumber\\
&&+ \int d^4\theta \frac{1}{k g^2} {\rm Tr} \Biggl[
  + \Phi^\dagger e^{-V} \Phi  e^{V}
  + {\Phi^c}^\dagger e^{-V} \Phi^c e^{V} 
\Biggr] \Biggr\}.
\label{eq:t2z6action}
\end{eqnarray}

From above action, we obtain  
the transformations of gauge multiplet under $\Omega$ as
\begin{eqnarray}
  V(\omega z, \omega^{n-1} \bar z) &=& (R_{\Omega})^{l_V}
 V(z, \bar z) (R_{\Omega}^{-1})^{m_V}~,~\,
\end{eqnarray}
\begin{eqnarray}
  \Sigma(\omega z, \omega^{n-1} \bar z) &=& \omega^{n-1}
(R_{\Omega})^{l_{\Sigma}} 
\Sigma(z, \bar z) (R_{\Omega}^{-1})^{m_{\Sigma}}~,~\,
\end{eqnarray}
\begin{eqnarray}
  \Phi(\omega z, \omega^{n-1} \bar z) &=& \omega^{n-1}
(R_{\Omega})^{l_{\Phi}} 
\Phi(z, \bar z) (R_{\Omega}^{-1})^{m_{\Phi}}~,~\,
\end{eqnarray}
\begin{eqnarray}
  \Phi^c(\omega z,  \omega^{n-1}\bar z) &=& \omega^{2}
(R_{\Omega})^{l_{\Phi^c}}  
\Phi^c(z, \bar z) (R_{\Omega}^{-1})^{m_{\Phi^c}}~,~\,
\end{eqnarray}
where $(l_V, m_V)$,  $(l_{\Sigma}, m_{\Sigma})$,
$(l_{\Phi}, m_{\Phi})$, and $(l_{\Phi^c}, m_{\Phi^c})$  
 are equal to $(1, 1)$  if the gauge fields were
in the representations ${\bf (8, 1, 1), (1, 8, 1), 
(1, 1, 8)}$, and
$(l_V, m_V)$,  $(l_{\Sigma}, m_{\Sigma})$,
$(l_{\Phi}, m_{\Phi})$, and $(l_{\Phi^c}, m_{\Phi^c})$
 are equal to
$(1, 0)$ if the gauge fields were
in the representation ${\bf (\bar 3, 3, 3)}$,
and $(l_V, m_V)$,  $(l_{\Sigma}, m_{\Sigma})$,
$(l_{\Phi}, m_{\Phi})$, and $(l_{\Phi^c}, m_{\Phi^c})$  
 are equal to
$(0, 1)$ if the gauge fields were
in the representation ${\bf (3, \bar 3, \bar 3)}$.

Moreover, we choose the following matrix representation for
$R_{\Omega}$
\begin{eqnarray}
R_{\Omega} = (+1, +1, +1)\otimes (\omega^{n_1}, \omega^{n_1},
\omega^{n_2})
\otimes (\omega^{n_3}, \omega^{n_4}, \omega^{n_5})~.~\,
\end{eqnarray}
In order to have the models with $SU(3)\times SU(2) \times U(1)^3$
gauge symmetry and 4-dimensional $N=1$ supersymmetry for the zero modes,
 we obtain the following constraints on $n_i$
\begin{eqnarray}
{\rm (a)~} 2 n_1+n_2 =0 ~{\rm mod}~n~,~\,
\end{eqnarray}
\begin{eqnarray}
{\rm (b)~} n_3+n_4+n_5 =0 ~{\rm mod}~n~,~\,
\end{eqnarray}
\begin{eqnarray}
{\rm (c)}~n_1\not= n_2 ~{\rm mod}~n~,~\,
\end{eqnarray} 
\begin{eqnarray}
{\rm (d)}~n_3\not= n_4 \not= n_5 ~{\rm mod}~n~,~\,
\end{eqnarray} 
\begin{eqnarray}
{\rm (e)~} |n_1-n_2| \not= 1 ~{\rm and}~ n-2 ~{\rm mod}~n~,~\,
\end{eqnarray}
\begin{eqnarray}
{\rm (f)~} |n_i-n_j| \not= 1 ~{\rm and}~ n-2 ~{\rm mod}~n,~{\rm for}~
i, j=3, 4, 5 ~{\rm and}~ i\not= j~,~\,
\end{eqnarray}
\begin{eqnarray}
{\rm (g)~} |n_i+n_j| \not= 0, 1 ~{\rm and}~ n-2 ~{\rm mod}~n,~{\rm for}~
i=1, 2 ~{\rm and}~ j=3, 4, 5~.~\,
\end{eqnarray}
Because $R_{\Omega} \subset
SU(3)_C\times SU(3)_L \times SU(3)_R$, we obtain the constraints
(a) and (b). And the constraints
 (c) and (d) will break the $SU(3)_L$ down to
$SU(2)_L\times U(1)$ and $SU(3)_R$ down to $U(1)^2$, respectively.
In addition, the constraints (e) and (f) will project out all the
zero modes of $\Sigma$, $\Phi$ and $\Phi^c$ in the 
representations ${\bf (8, 1, 1), (1, 8, 1), (1, 1, 8)}$,
and the constraint (g) will project out all the zero modes of 
$V$, $\Sigma$, $\Phi$ and $\Phi^c$ in the representations 
${\bf (\bar 3, 3, 3)}$ and 
${\bf (3, \bar 3, \bar 3)}$.

Let us give the simplest model with $Z_9$ symmetry, the matrix
representation for
$R_{\Omega}$ is
\begin{eqnarray}
R_{\Omega} = (+1, +1, +1)\otimes (\omega^{2}, \omega^{2}, \omega^{5})
\otimes (+1, \omega^{3}, \omega^{6})~.~\,
\end{eqnarray}
It is easy to check that all the constraints are satisfied.

First, we consider that the extra space manifold is the annulus $A^2$. 
For the zero modes, we have 4-dimensional $N=1$ supersymmetry and
$SU(3)\times SU(2) \times U(1)^3$ gauge symmetry in the bulk and on
the 4-branes at $r=R_1$ and $r=R_2$. Including the
KK states, we will have the 4-dimensional $N=4$ supersymmetry and
 $E_6$ gauge symmetry in the bulk, and on
the 4-branes at $r=R_1$ and $r=R_2$. 

Second, we consider that the extra space manifold is
the disc $D^2$. For the zero modes, 
we have 4-dimensional $N=1$ supersymmetry and
$SU(3)\times SU(2) \times U(1)^3$ gauge symmetry in the bulk and on
the 4-brane at $r=R_2$. Including all the
KK states, we will have the 4-dimensional $N=4$ supersymmetry and
 $E_6$ gauge symmetry in the bulk, and on
the 4-brane at $r=R_2$. In addition, because the origin 
($r=0$) is the fixed point under the $Z_9$ symmetry,
 we always have the 4-dimensional $N=1$ supersymmetry and
$SU(3)\times SU(2) \times U(1)^3$ gauge symmetry on the 3-brane at origin
in which only the zero modes exist.
 And if we put the Standard Model fermions on the 3-brane at origin,
 the extra dimensions
can be large and the gauge hierarchy problem can be solved 
for there does not exist the proton decay problem at all.

\section{Discussion and Conclusion}
The extra gauge symmetry must be broken around or above
 the TeV scale. This can be done by Higgs mechanism, for
example, the $SU(4)$ can be broken down to $SU(3)_C$ and 
$SU(2)_R$ can be broken by introducing
the Higgs fields in their fundamental representations,
and the extra $U(1)$ can be broken by introducing the Standard Model
singlets which are charged under the extra $U(1)$. The right-handed
neutrinos
might also become massive during the extra gauge symmetry breaking. 
And with the ansatz that there exist discrete symmetries in the 
 neighborhoods of the branes, one can discuss the general 
$E_6$ breaking on the space-time $M^4\times M^1\times M^1$,
where the extra dimensions can be large and
the KK states can be set arbitrarily heavy~\cite{JUNII}.
On the outlook, the gauge coupling unification, supersymmetry breaking,
the $\mu$ problem,
 how to forbid the proton decay operators by $R$ symmetry, and
how to explain the fermion mass hierarchy and mixing angles in our
models deserve further study.

In short, we have studied the $N=2$ supersymmetric $E_6$ models on the 
6-dimensional space-time where the supersymmetry and 
gauge symmetry can be broken by the discrete symmetry. 
On the space-time 
$M^4\times S^1/(Z_2\times Z_2') \times S^1/(Z_2\times Z_2')$,
for the zero modes, we obtain the 4-dimensional $N=1$
supersymmetric models with gauge groups 
$SU(3)\times SU(2) \times SU(2) \times U(1)^2$, 
$SU(4)\times SU(2) \times SU(2) \times U(1)$, and
$SU(3)\times SU(2) \times U(1)^3$ with one extra pair of Higgs doublets
from the vector multiplet.
In addition, considering that
the extra space manifold is the annulus $A^2$ and disc $D^2$,
we list all the constraints on constructing the 4-dimensional $N=1$
supersymmetric $SU(3)\times SU(2) \times U(1)^3$ models for the 
zero modes, and give the simplest model with $Z_9$ symmetry.
 
\section*{Acknowledgments}
T. Li would like to thank Institute of Theoretical Physics, Academia
Sinica,
P. R. China for hospitality.
This work was supported in part by the Natural Science
Foundation of China and by the U.S.~Department of Energy under Grant 
 No.~DOE-EY-76-02-3071 and W-31-109-ENG-38.

\end{document}